\documentclass[aps,prd,reprint,nofootinbib]{revtex4-1}

\usepackage{amsmath,amssymb,amsfonts}
\usepackage{amsthm}
\usepackage[utf8]{inputenc}
\usepackage{graphicx}
\usepackage{url}
\usepackage{hyperref}
\usepackage{bm,bbm}
\usepackage{multirow}

\newcommand{\nwse}[3]{\ensuremath{#1^{#2}_{\phantom{#2} #3}}}

\newcommand{\Tr}{\ensuremath{\text{Tr}}}
\newcommand{\um}{\ensuremath{\mathbbm{1}}}

\newcommand{\ucsc}{Departamento de Matemática y Física Aplicadas, Universidad Católica de la Santísima Concepción, Alonso de Ribera 2850, 4090541 Concepción, Chile}

\begin{document}

\title{Dirac Matrices for Chern--Simons Gravity}

\author{Fernando Izaurieta}
\email{fizaurie@ucsc.cl}

\author{Ricardo Ram\'{\i}rez}
\email{ricramirez@ucsc.cl}

\author{Eduardo Rodr\'{\i}guez}
\email{edurodriguez@ucsc.cl}
\affiliation{\ucsc}

\date{\today}

\begin{abstract}
A genuine gauge theory for the Poincaré, de~Sitter or anti-de~Sitter algebras can be constructed in $\left( 2n-1 \right)$-dimensional spacetime by means of the Chern--Simons form, yielding a gravitational theory that differs from General Relativity but shares many of its properties, such as second order field equations for the metric.
The particular form of the Lagrangian is determined by a rank $n$, symmetric tensor invariant under the relevant algebra.
In practice, the calculation of this invariant tensor can be reduced to the computation of the trace of the symmetrized product of $n$ Dirac Gamma matrices $\Gamma_{ab}$ in $2n$-dimensional spacetime.
While straightforward in principle, this calculation can become extremely cumbersome in practice.
For large enough $n$, existing computer algebra packages take an inordinate long time to produce the answer or plainly fail having used up all available memory.
In this talk we show that the general formula for the trace of the symmetrized product of $2n$ Gamma matrices $\Gamma_{ab}$ can be written as a certain sum over the integer partitions $s$ of $n$, with every term being multiplied by a numerical coefficient $\alpha_{s}$.
We then give a general algorithm that computes the $\alpha$-coefficients as the solution of a linear system of equations generated by evaluating the general formula for different sets of tensors $B^{ab}$ with random numerical entries.
A recurrence relation between different coefficients is shown to hold and is used in a second, ``minimal'' algorithm to greatly speed up the computations.
Runtime of the minimal algorithm stays below 1~min on a typical desktop computer for up to $n=25$, which easily covers all foreseeable applications of the trace formula.
\end{abstract}

\maketitle

\section{Introduction}
\label{sec:intro}

There's more to higher-dimensional gravity than Einstein and Hilbert~\cite{Lan38,Lov71,Des82a,Des82b,Zwi85,Zum85,Mar91}.

Chern--Simons (CS) gravity in $d = 2n-1$ dimensions is a gauge theory for the Poincaré, de~Sitter or anti-de~Sitter (AdS) algebras, depending on the value of the cosmological constant~\cite{Cha89,Banh93}.

Let us focus on the AdS algebra, $\mathfrak{so} \left( d-1,2 \right)$.
A convenient matrix representation is provided by
$\Gamma_{AB} = \Gamma_{[A} \Gamma_{B]}$, where
$\Gamma_{A}$ are Dirac matrices in $D = d+1 = 2n$ dimensions:%
\footnote{The indices run as follows: $A,B=0,1,\ldots,D-1$, $a,b=0,1,\ldots,d-1$.}
\begin{align}
\bm{J}_{ab} & = \frac{1}{2} \Gamma_{ab}, \\
\bm{P}_{a} & = \frac{1}{2} \Gamma_{a,d}.
\end{align}

The Lagrangian for CS gravity is shaped to a great extent by a rank-$n$,
AdS-invariant symmetric polynomial $\left\langle \cdots \right\rangle$.

This polynomial can be identified with any of the following traces:
\begin{itemize}
 \item
 $\Tr \left\{ \Gamma_{A_{1} B_{1}} \cdots \Gamma_{A_{n} B_{n}} \right\}$
 (Lorentz scalar)
 \item
 $\Tr \left( \Gamma_{\ast} \left\{ \Gamma_{A_{1} B_{1}} \cdots \Gamma_{A_{n} B_{n}} \right\} \right)$
 (Lorentz pseudoscalar),
\end{itemize}
where $\{ \cdots \}$ denotes symmetrized matrix product.

The pseudoscalar trace reads
\begin{equation}
\Tr \left( \Gamma_{\ast} \left\{ \Gamma_{A_{1} B_{1}} \cdots \Gamma_{A_{n} B_{n}} \right\} \right) =
\gamma
\epsilon_{A_{1} B_{1} \cdots A_{n} B_{n}},
\end{equation}
where $\gamma$ is a numerical coefficient.
Use of this invariant polynomial brings in the Lanczos--Lovelock~\cite{Lov71,Banh93} family of Lagrangians into CS gravity.

The scalar trace, on the other hand, is more involved.

In this work we provide two algorithms that can be used to efficiently compute the scalar trace for any $n$ and for any spacetime dimension $d$ (without any implied relation between $n$ and $d$).

\section{Formulation of the Problem and Results}
\label{sec:formulation}

Let us consider Dirac matrices $\Gamma_{a}$, $a = 0, \ldots, d-1$, in $d$-dimensional Minkowski spacetime.
By definition, they satisfy the Clifford algebra~\cite{Fre86}
\begin{equation}
\Gamma_{a} \Gamma_{b} + \Gamma_{b} \Gamma_{a} = 2 \eta_{ab} \um,
\label{eq:Gdef}
\end{equation}
where $\eta_{ab} = \left( - + \cdots + \right)$ is the usual Minkowski metric and
\um\ stands for the $m \times m$ unit matrix, with $m = 2^{\left\lfloor d/2 \right\rfloor}$.

The $\Gamma$-matrices which are the subject of this work are defined as
\begin{equation}
\Gamma_{ab} =
\Gamma_{[a} \Gamma_{b]} =
\frac{1}{2} \left( \Gamma_{a} \Gamma_{b} - \Gamma_{b} \Gamma_{a} \right).
\label{eq:G2}
\end{equation}

For completeness, let us define the symmetrized product of $n$ matrices $M_{i}$, $i=1,\ldots,n$, as
\begin{equation}
\left\{ M_{1} \cdots M_{n} \right\} =
\frac{1}{n!}
\sum_{\pi \in S_{n}}
M_{\pi \left( 1 \right)}
\cdots
M_{\pi \left( n \right)},
\end{equation}
where the sum extends over all permutations $\pi$ in the symmetric group $S_{n}$.

Experience shows that the trace is most efficiently written with all matrices multiplied by arbitrary antisymmetric tensors.
Take, for instance, the trace of the symmetrized product of two Gamma matrices, and compare the following equations:
\begin{align}
\Tr \left\{ \Gamma_{ab} \Gamma_{cd} \right\} & =
m \left( \eta_{ad} \eta_{bc} - \eta_{ac} \eta_{bd} \right),
\label{eq:T2} \\
A^{ab} B^{cd}
\Tr \left\{ \Gamma_{ab} \Gamma_{cd} \right\} & =
2m \nwse{A}{a}{b} \nwse{B}{b}{a}.
\label{eq:T2AA}
\end{align}
The two terms on the right-hand side of eq.~(\ref{eq:T2}) have collapsed into one in eq.~(\ref{eq:T2AA}).
Greater simplifications are achieved for more complicated cases.
If desired, eq.~(\ref{eq:T2}) can be recovered from  eq.~(\ref{eq:T2AA}) by means of the formal replacement
$A^{ab} \rightarrow \delta^{ab}_{cd}$,
$B^{ab} \rightarrow \delta^{ab}_{cd}$,
where $\delta^{ab}_{cd}$ is the generalized Kronecker delta.

Let $B_{i}^{ab}$, $i=1,2,3,\ldots$, be arbitrary antisymmetric tensors, and let us define
\begin{equation}
\beta_{i} = B_{i}^{ab} \Gamma_{ab}.
\label{eq:betadef}
\end{equation}

The symmetrized product of $n$ $\beta$-matrices can be written as a linear combination of
matrices $\Gamma_{a_{1} \cdots a_{p}} = \Gamma_{[a_{1}} \cdots \Gamma_{a_{p}]}$,
with
$p = 0, 4, 8, \ldots, 2n$ (for $n$ even) or
$p = 2, 6, 10, \ldots, 2n$ (for $n$ odd).
The only term that contributes to the trace is that proportional to the identity matrix ($p=0$).
For odd $d$, however, the $\Gamma_{a_{1} \cdots a_{d}}$ matrix is also proportional to the identity
and must be generically taken into account when computing the trace.
The expansion of the symmetrized product of the $\beta$-matrices includes only $\Gamma$-matrices with an even number of indices, so that the $\Gamma_{a_{1} \cdots a_{d}}$-term never actually shows up in our case.
In particular, this means that the trace of the symmetrized product of an odd number of $\beta$-matrices vanishes identically.

The trace of the symmetrized product of $2n$ $\beta$-matrices, on the other hand, can be written as
\begin{equation}
\Tr \left\lbrace \beta_{1} \cdots \beta_{2n} \right\rbrace =
m \sum_{s \vdash n}
\alpha_{s} \mathcal{B}^{\left( s \right)},
\label{eq:general}
\end{equation}
where the notation $s \vdash n$~\cite{And98}
indicates that the sum must be performed over all integer partitions $s$ of $n$,
and $\mathcal{B}^{\left( s \right)}$ stands for the following sum of contractions of $B$-tensors:
\begin{equation}
\mathcal{B}^{\left( s \right)} =
\sum_{\left\langle i_{1} \cdots i_{2n} \right\rangle}
\prod_{j=1}^{r}
\left\langle
 B_{i_{2 s_{1} + \cdots + 2 s_{j-1} + 1}}
 \cdots
 B_{i_{2 s_{1} + \cdots + 2 s_{j}}}
\right\rangle.
\label{eq:B}
\end{equation}
In eq.~(\ref{eq:B}), the notation
$\left\langle i_{1} \cdots i_{2n} \right\rangle$
is used to indicate that the sum must be performed over all
$i_{1}, \ldots, i_{2n} \in \left\{ 1, \ldots, 2n \right\}$,
with the restriction that they be all different.
This implements the permutation of all $\beta$-matrices.
Every term in the sum contains the product of $r$ factors of the form
$\left\langle B_{1} \cdots B_{q} \right\rangle$,
where $r$ is the length of the partition
$s = \left( s_{1}, \ldots, s_{r} \right)$, with
$n = s_{1} + \cdots + s_{r}$.
The $j$-th factor in the product represents the trace of the product of $2s_{j}$ $B$-tensors, i.e.,
\begin{equation}
\left\langle B_{1} \cdots B_{q} \right\rangle =
\nwse{\left( B_{1} \right)}{c_{1}}{c_{2}}
\nwse{\left( B_{2} \right)}{c_{2}}{c_{3}}
\cdots
\nwse{\left( B_{q} \right)}{c_{q}}{c_{1}},
\label{eq:Bq}
\end{equation}
with $q = 2s_{j}$.

To every term in eq.~(\ref{eq:general}), i.e., to every partition $s$ of $n$, there corresponds an $\alpha_{s}$ coefficient.
Numerical values for the $\alpha$-coefficients corresponding to the partitions of $n = 1, \ldots, 7$ are given in Table~\ref{tab:alpha17}.

\begin{table}[tphb]
\caption{\label{tab:alpha17}$\alpha$-coefficients corresponding to the partitions of $n = 1, \ldots, 7$.}
\begin{ruledtabular}
\begin{tabular}{clr}
$n$ & $s$ & $\alpha_{s}$ \\
\hline
1 & 1 & 1 \\
\hline
2 & $1+1$ & $1/2$ \\
  & 2 & $-2/3$ \\
\hline
3 & $1+1+1$ & $1/6$ \\
  & $2+1$ & $-2/3$ \\
  & 3 & $32/45$ \\
\hline
4 & $1+1+1+1$ & $1/24$ \\
  & $2+1+1$ & $-1/3$ \\
  & $2+2$ & $2/9$ \\
  & $3+1$ & $32/45$ \\
  & 4 & $-272/315$ \\
\hline
5 & $1+1+1+1+1$ & $1/120$ \\
  & $2+1+1+1$ & $-1/9$ \\
  & $2+2+1$ & $2/9$ \\
  & $3+1+1$ & $16/45$ \\
  & $3+2$ & $-64/135$ \\
  & $4+1$ & $-272/315$ \\
  & 5 & $15872/14175$ \\
\hline
6 & $1+1+1+1+1+1$ & $1/720$ \\
  & $2+1+1+1+1$ & $-1/36$ \\
  & $2+2+1+1$ & $1/9$ \\
  & $2+2+2$ & $-4/81$ \\
  & $3+1+1+1$ & $16/135$ \\
  & $3+2+1$ & $-64/135$ \\
  & $3+3$ & $512/2025$ \\
  & $4+1+1$ & $-136/315$ \\
  & $4+2$ & $544/945$ \\
  & $5+1$ & $15872/14175$ \\
  & 6 & $-707584/467775$ \\
\hline
7 & $1+1+1+1+1+1+1$ & $1/5040$ \\
  & $2+1+1+1+1+1$ & $-1/180$ \\
  & $2+2+1+1+1$ & $1/27$ \\
  & $2+2+2+1$ & $-4/81$ \\
  & $3+1+1+1+1$ & $4/135$ \\
  & $3+2+1+1$ & $-32/135$ \\
  & $3+2+2$ & $64/405$ \\
  & $3+3+1$ & $512/2025$ \\
  & $4+1+1+1$ & $-136/945$ \\
  & $4+2+1$ & $544/945$ \\
  & $4+3$ & $-8704/14175$ \\
  & $5+1+1$ & $7936/14175$ \\
  & $5+2$ & $-31744/42525$ \\
  & $6+1$ & $-707584/467775$ \\
  & 7 & $89473024/42567525$ \\
\end{tabular}
\end{ruledtabular}
\end{table}

The following examples for $n = 1, \ldots, 4$ should help clarify the meaning of eqs.~(\ref{eq:general}) and~(\ref{eq:B}):
\begin{equation}
\Tr
\left\lbrace
 \beta_{1} \beta_{2}
\right\rbrace
= m
\sum_{\left\langle ij \right\rangle}
 \alpha_{1}
 \left\langle B_{i} B_{j} \right\rangle,
\end{equation}
\begin{align}
\Tr
\left\lbrace
 \beta_{1} \cdots \beta_{4}
\right\rbrace
& = m
\sum_{\left\langle ijkl \right\rangle}
 \left[
  \alpha_{2}
  \left\langle B_{i} B_{j} B_{k} B_{l} \right\rangle +
 \right.
 \nonumber \\ &
 \left. +
  \alpha_{11}
  \left\langle B_{i} B_{j} \right\rangle
  \left\langle B_{k} B_{l} \right\rangle
 \right],
\end{align}
\begin{align}
\Tr
\left\lbrace
 \beta_{1} \cdots \beta_{6}
\right\rbrace
& = m
\sum_{\left\langle i_{1} \cdots i_{6} \right\rangle}
 \left[
  \alpha_{3}
  \left\langle B_{i_{1}} \cdots B_{i_{6}} \right\rangle +
 \right.
 \nonumber \\ & +
  \alpha_{21}
  \left\langle B_{i_{1}} \cdots B_{i_{4}} \right\rangle
  \left\langle B_{i_{5}} B_{i_{6}} \right\rangle +
 \nonumber \\ &
 \left. +
  \alpha_{111}
  \left\langle B_{i_{1}} B_{i_{2}} \right\rangle
  \left\langle B_{i_{3}} B_{i_{4}} \right\rangle
  \left\langle B_{i_{5}} B_{i_{6}} \right\rangle
 \right],
\label{eq:n=3}
\end{align}
\begin{align}
\Tr
\left\lbrace
 \beta_{1} \cdots \beta_{8}
\right\rbrace
& = m
\sum_{\left\langle i_{1} \cdots i_{8} \right\rangle}
 \left[
  \alpha_{4}
  \left\langle B_{i_{1}} \cdots B_{i_{8}} \right\rangle +
 \right.
 \nonumber \\ & +
  \alpha_{31}
  \left\langle B_{i_{1}} \cdots B_{i_{6}} \right\rangle
  \left\langle B_{i_{7}} B_{i_{8}} \right\rangle +
 \nonumber \\ & +
  \alpha_{22}
  \left\langle B_{i_{1}} \cdots B_{i_{4}} \right\rangle
  \left\langle B_{i_{5}} \cdots B_{i_{8}} \right\rangle +
 \nonumber \\ & +
  \alpha_{211}
  \left\langle B_{i_{1}} \cdots B_{i_{4}} \right\rangle
  \left\langle B_{i_{5}} B_{i_{6}} \right\rangle
  \left\langle B_{i_{7}} B_{i_{8}} \right\rangle +
 \nonumber \\ &
 \left. +
  \alpha_{1111}
  \left\langle B_{i_{1}} B_{i_{2}} \right\rangle
  \left\langle B_{i_{3}} B_{i_{4}} \right\rangle
  \left\langle B_{i_{5}} B_{i_{6}} \right\rangle
  \left\langle B_{i_{7}} B_{i_{8}} \right\rangle
 \right].
\end{align}

The proof of eq.~(\ref{eq:general}) is by exhaustion;
the right-hand side includes all possible terms that may contribute to the trace of the symmetrized product of $2n$ $\beta$-matrices.%
\footnote{The formula for $\Tr \left( \Gamma_{\ast} \left\{ \beta_{1} \cdots \beta_{n} \right\} \right)$ includes pseudoscalar terms that appear in certain dimensions $d$ (e.g., $\epsilon_{abcd} B_{i}^{ab} B_{j}^{cd}$ for $d=4$) but are absent from $\Tr \left\{ \beta_{1} \cdots \beta_{2n} \right\}$, where only Lorentz scalars are allowed. Here $\Gamma_{\ast} = \Gamma_{0} \cdots \Gamma_{d-1}$ is the $d$-dimensional generalization of $\gamma_{5}$ in $d=4$.}

Our approach to the computation of the $\alpha$-coefficients is the subject of section~\ref{sec:method}.

\section{Method}
\label{sec:method}

\subsection{General Algorithm}

The central observation behind the algorithm used in the computation of the $\alpha$-coefficients
shown in Table~\ref{tab:alpha17}
is the fact that eq.~(\ref{eq:general}) is valid for \emph{arbitrary} tensors $B_{i}^{ab}$.

For illustration purposes, let us focus first on the $n=3$ case.
Eq.~(\ref{eq:n=3}) simplifies greatly if we choose all $B$-tensors to be equal,
since in this case the sum over all different permutations of
$i_{1}, \ldots, i_{6} \in \left\{ 1, \ldots, 6 \right\}$
is trivially performed. The result reads
\begin{align}
\frac{1}{6!m} \Tr \left( \beta^{6} \right)  =
\alpha_{3} \left\langle B^{6} \right\rangle +
\alpha_{21} \left\langle B^{4} \right\rangle \left\langle B^{2} \right\rangle +
\alpha_{111} \left\langle B^{2} \right\rangle^{3}.
\label{eq:n=3B}
\end{align}

We wish to cast eq.~(\ref{eq:n=3B}) as a linear equation with three unknowns,
namely, $\alpha_{3}$, $\alpha_{21}$ and $\alpha_{111}$.
To do this we need to be able to assign numerical values to the left-hand side
and to the various $\left\langle B^{q} \right\rangle$-terms that appear on the right-hand side.
We accomplish this by
(i)~picking some antisymmetric tensor $B^{ab}$ with random numerical entries and
(ii)~choosing an explicit representation for the $\Gamma$-matrices.%
\footnote{See section~\ref{sec:final} for a discussion of the choice of spacetime dimension $d$ in which to carry out the computation.}
We emphasize that the possibility of choosing the $B$-tensors at will relies upon the fact that eq.~(\ref{eq:general}) is valid for arbitrary $B_{i}$'s.

To be able to solve for the $\alpha$-coefficients we need two more equations.
These are readily obtained by randomly selecting two further $B$-tensors.
Denoting the three different choices for the $B$-tensors by $B_{k}$, with $k = 1,2,3$,
we obtain the following $3 \times 3$ linear system:
\begin{align}
Z_{1}^{\left( 111 \right)} \alpha_{111} +
Z_{1}^{\left( 21 \right)} \alpha_{21} +
Z_{1}^{\left( 3 \right)} \alpha_{3} & =
T_{1}, \\
Z_{2}^{\left( 111 \right)} \alpha_{111} +
Z_{2}^{\left( 21 \right)} \alpha_{21} +
Z_{2}^{\left( 3 \right)} \alpha_{3} & =
T_{2}, \\
Z_{3}^{\left( 111 \right)} \alpha_{111} +
Z_{3}^{\left( 21 \right)} \alpha_{21} +
Z_{3}^{\left( 3 \right)} \alpha_{3} & =
T_{3},
\end{align}
where
\begin{align}
T_{k} & = \frac{1}{6!m} \Tr \left( \beta_{k}^{6} \right), \\
Z_{k}^{\left( 111 \right)} & = \left\langle B_{k}^{2} \right\rangle^{3}, \\
Z_{k}^{\left( 21 \right)} & = \left\langle B_{k}^{4} \right\rangle \left\langle B_{k}^{2} \right\rangle, \\
Z_{k}^{\left( 3 \right)} & = \left\langle B_{k}^{6} \right\rangle.
\end{align}

The method to compute the $\alpha$-coefficients for any value of $n$ is now clear
and can be summarized in the following sequence:
\begin{enumerate}
\item Let $p = p \left( n \right)$ be the number of partitions of $n$.%
\footnote{The function $p \left( n \right)$ is called the ``partition function'' in the mathematical literature~\cite{And98}.}
\item Choose an explicit representation for the $\Gamma$-matrices (see, e.g., Ref.~\cite{VanPro99}).
\item For $k = 1, \ldots, p$, do:
 \begin{enumerate}
 \item Pick an antisymmetric tensor $B_{k}^{ab}$ with random numerical entries.
 \item Compute
 \begin{equation}
 T_{k} = \frac{1}{\left( 2n \right) ! m}
 \Tr \left( \beta_{k}^{2n} \right),
 \end{equation}
 where $\beta_{k} = B_{k}^{ab} \Gamma_{ab}$.
 \item For every partition $s \vdash n$, with $n = s_{1} + \cdots + s_{r}$, compute
 \begin{equation}
 Z_{k}^{\left( s \right)} = \prod_{j=1}^{r} \left\langle B_{k}^{2s_{j}} \right\rangle.
 \end{equation}
 The notation $\left\langle B_{k}^{q} \right\rangle$ stands for [see eq.~(\ref{eq:Bq})]
 \begin{equation}
 \left\langle B_{k}^{q} \right\rangle =
 \nwse{\left( B_{k} \right)}{c_{1}}{c_{2}}
 \nwse{\left( B_{k} \right)}{c_{2}}{c_{3}}
 \cdots
 \nwse{\left( B_{k} \right)}{c_{q}}{c_{1}}.
 \end{equation}
 \end{enumerate}
\item The $\alpha$-coefficients are the solution to the $p \times p$ linear system of equations
\begin{equation}
\sum_{s \vdash n} Z_{k}^{\left( s \right)} \alpha_{s} = T_{k}
\qquad
\left( k = 1, \ldots, p \right).
\label{eq:T=Zx}
\end{equation}
\end{enumerate}

\subsection{Minimal Algorithm}

Careful inspection of the $\alpha$-coefficients shown in Table~\ref{tab:alpha17} shows that there exists a recurrence relation among different coefficients.

Let $s$ be a partition of $n$.
The frequency representa\-tion \cite{And98} of $s$ is the notation
$s = \left( 1^{\mu_{1}} \ 2^{\mu_{2}} \ \cdots \right)$,
where $\mu_{j}$ represents the multiplicity of $j$,
i.e., the number of times that a given integer $j$ appears in $s$.

We find that the coefficient $\alpha_{s}$ corresponding to the partition
$s = \left( 1^{\mu_{1}} \ 2^{\mu_{2}} \ \cdots \right)$ can be written as
\begin{equation}
\alpha_{s} = \prod_{j=1}^{n} \frac{\alpha_{j}^{\mu_{j}}}{\mu_{j}!},
\label{eq:rr}
\end{equation}
where $\alpha_{j}$ are the coefficients associated with the ``elementary'' partitions
$1=1$, $2=2$, $3=3$, etc.

For example, all coefficients associated with the non-elementary partitions of $n = 1,2,3$
can be computed from $\alpha_{1}$, $\alpha_{2}$ and $\alpha_{3}$ by means of the equations
\begin{align}
\alpha_{11} & =
\frac{\alpha_{1}^{2}}{2!}
\frac{\alpha_{2}^{0}}{0!} =
\frac{1}{2}, \\
\alpha_{111} & =
\frac{\alpha_{1}^{3}}{3!}
\frac{\alpha_{2}^{0}}{0!}
\frac{\alpha_{3}^{0}}{0!} =
\frac{1}{6}, \\
\alpha_{21} & =
\frac{\alpha_{1}^{1}}{1!}
\frac{\alpha_{2}^{1}}{1!}
\frac{\alpha_{3}^{0}}{0!} =
-\frac{2}{3}.
\end{align}
Of course, this recurrence relation also holds for more complicated cases, such as
\begin{equation}
\alpha_{3211} =
\frac{\alpha_{1}^{2}}{2!}
\frac{\alpha_{2}^{1}}{1!}
\frac{\alpha_{3}^{1}}{1!}
\frac{\alpha_{4}^{0}}{0!}
\frac{\alpha_{5}^{0}}{0!}
\frac{\alpha_{6}^{0}}{0!}
\frac{\alpha_{7}^{0}}{0!}
= -\frac{32}{135}.
\end{equation}
When applied to an elementary coefficient, eq.~(\ref{eq:rr}) yields an identity.

The recurrence relation in eq.~(\ref{eq:rr}) can be used to compute the values for the $\alpha$-coefficients associated with all the non-elementary partitions of $n$.
Its use, however, requires knowledge of the elementary coefficients, for which no closed formula is available.
This situation suggests a ``minimal'' algorithm that
(i)~calculates elementary coefficients in a manner analogous to that of the ``general'' algorithm and
(ii)~computes non-elementary coefficients from eq.~(\ref{eq:rr}).

The following sequence describes such an algorithm:
\begin{enumerate}
\item Let $N$ be the maximum integer for which we wish to calculate the $\alpha$-coefficients.
\item Choose an explicit representation for the $\Gamma$-matrices.
\item Pick an antisymmetric tensor $B^{ab}$ with random numerical entries.%
\footnote{We took $d=2$ and $B^{01} = +1$, since a two-index antisymmetric tensor has only one degree of freedom in two spacetime dimensions, and overall numerical factors are not significant for the calculation. See section~\ref{sec:final} for a discussion of the choice of spacetime dimension $d$ in which to carry out the computation.}
\item For $n = 1, \ldots, N$, do:
 \begin{enumerate}
 \item Compute
 \begin{equation}
 T = \frac{1}{\left( 2n \right) ! m}
 \Tr \left( \beta^{2n} \right),
 \end{equation}
 where $\beta = B^{ab} \Gamma_{ab}$.
 \item For every partition $s \vdash n$, with $n = s_{1} + \cdots + s_{r}$, compute
 \begin{equation}
 Z^{\left( s \right)} = \prod_{j=1}^{r} \left\langle B^{2s_{j}} \right\rangle.
 \end{equation}
 \item Use the recurrence relation~(\ref{eq:rr}) to calculate all non-elementary coefficients associated with the partitions of $n$ (this step is empty for $n=1$).
 \item Solve
 \begin{equation}
 \sum_{s \vdash n} Z^{\left( s \right)} \alpha_{s} = T
 \end{equation}
 for $\alpha_{n}$ (this is a linear equation with one unknown).
 \end{enumerate}
\end{enumerate}

\section{Discussion and Conclusions}
\label{sec:final}

The algorithms described in section~\ref{sec:method} turn around the problem of finding formulas for the trace of a product of Gamma matrices.
The usual textbook approach starts with eq.~(\ref{eq:Gdef}) and \emph{deduces} the required formulas from there.
Our approach here works the other way around.
We start by identifying the general form of the equation for the trace of the symmetrized product of $2n$ $\beta$-matrices.
Eq.~(\ref{eq:general}) amounts to such an identification, since it contains all possible sums of $B$-contractions that may contribute to the trace. The $\alpha$-coefficients appear as undetermined parameters, which are computed by demanding validity of eq.~(\ref{eq:general}) in several nontrivial cases.

As stressed in section~\ref{sec:method}, our method works because eq.~(\ref{eq:general}) holds for \emph{arbitrary} antisymmetric tensors $B_{i}^{ab}$.
We have used $B$-tensors with random numerical entries to generate the linear system of equations whose solution provides the $\alpha$-coef\-ficients.
In this sense our approach bears some resemblance to Monte Carlo methods, where random numbers play a crucial role.
The use of random matrices,%
\footnote{To be precise, what we use are actually two-index antisymmetric tensors with random numerical entries.}
however, is not essential to our calculation.
All that is required for the general algorithm to succeed is a set of $B$-tensors such that every iteration produces an equation for the $\alpha$-coefficients that is linearly independent from the rest, yielding a full-rank $Z$ matrix [cf.~eq.~(\ref{eq:T=Zx})].

The solution we find is, of course, independent of the choice of $B$-tensors; this is conceptually clear, but can also be verified by running the algorithm several times with different sets of (randomly generated) $B$-tensors.
The fact that the same solution is obtained every time confirms both this independence and the correctness of eq.~(\ref{eq:general}),
i.e., that no other terms can be added to the trace.

The $\alpha$-coefficients are also independent of the spacetime dimension $d$, which means that the algorithm should in principle work for any $d$ we choose.
There is, however, an important caveat.
To produce a solvable system one needs the $B$-tensors to have a sufficient number of independent components, so that the successive iterations of the algorithm yield linearly independent equations.
We find that there is a minimum spacetime dimension $d=2n$ that allows the $Z$ matrix to achieve full rank.
This means that the general algorithm must be run with $d \geq 2n$ in order for a solution to be produced.

The minimal algorithm, with only one linear equation to be solved, works even with a minimum spacetime dimension of $d=2$.

Is our approach any better than the textbook method?
One way to probe into this question is to compare the runtime of both.
The textbook method can be implemented in, e.g.,
Kasper Peeters' excellent computer algebra system ``Ca\-dabra''~\cite{Pee06,Pee07}.
We were able to deduce, starting only from the definition of Dirac matrices, the $\alpha$-coefficients for $n = 1,2,3$.
The $n=3$ case took some 30~min to be solved on a typical desktop computer,%
\footnote{In 2011 this meant a 3.20-GHz CPU, with 3.7~GB of memory.}
while the $n=4$ case caused the program to crash.
This approach, of course, requires hardly any input and produces the full sought-after formula.
Starting from eq.~(\ref{eq:general}), we programmed our general algorithm in the computer algebra system ``Maxima''~\cite{maxima} and were able to run it successfully for $n = 1, \ldots, 7$.
The $n=8$ case caused Maxima to run out of memory.
Runtime for $n = 1, \ldots, 4$ was negligible, while the $n=7$ case took under half an hour.
The minimal algorithm, which we also programmed in Maxima, had negligible runtime even for $N=25$.
Table~\ref{tab:runtime} summarizes runtime for these different scenarios.

Complexity for the general algorithm grows exponentially with $n$.
Complexity for the minimal algorithm, on the other hand, grows linearly with $p$.%
\footnote{An asymptotic approximation for the partition function $p \left( n \right)$ is given by the Hardy--Ramanujan equation, $p \left( n \right) = \left( 1/4n\sqrt{3} \right) \exp \left( \pi \sqrt{2n/3} \right)$~\cite{And98}.}
All foreseeable applications of the formula for the trace of a product of $2n$ Gamma matrices are well covered by the minimal algorithm with negligible runtime.

\begin{table}[thbp]
 \caption{\label{tab:runtime}Approximate runtime for the textbook method (as implemented in Cadabra) and the general and minimal algorithms (as implemented in Maxima) on a typical desktop computer. For the minimal algorithm, the first column is understood to mean $N$, the maximum integer for which the $\alpha$-coefficients are computed. The second column lists the partition function of $n$, which corresponds to the number of $\alpha$-coefficients to be determined.}
 \begin{ruledtabular}
 \begin{tabular}{ccccc}
 \multirow{2}{*}{$n$} & \multirow{2}{*}{$p$} & Textbook & General   & Minimal \\
   &  & Method   & Algorithm & Algorithm \\
\hline
  1 & 1  & negligible    & negligible    & negligible \\
  2 & 2  & negligible    & negligible    & negligible \\
  3 & 3  & $\sim 30$~min & negligible    & negligible \\
  4 & 5  & crashed       & negligible    & negligible \\
  5 & 7  &               & negligible    & negligible \\
  6 & 11 &               & $\sim 1$~min  & negligible \\
  7 & 15 &               & $\sim 24$~min & negligible \\
  8 & 22 &               & crashed       & negligible \\
  9 & 30 &               &               & negligible \\
  \vdots & \vdots &      &               & \vdots     \\
  26 & 2436 &            &               & $\sim 1$~min   \\
  28 & 3718 &            &               & $\sim 2$~min   \\
  30 & 5604 &            &               & $\sim 5$~min   \\
 \end{tabular}
 \end{ruledtabular}
\end{table}

\begin{acknowledgments}
The authors wish to thank
Tomás Barrios and
Ruth Sandoval
for many friendly, helpful and enlightening conversations on the subject of this work.
F.~I.\ and E.~R.\ were supported by the
National Commission for Scientific \& Technological Research, Chile,
through Fondecyt research grants 11080200 and 11080156, respectively.
\end{acknowledgments}

\bibliographystyle{utphys}
\bibliography{biblio2012}

\end{document}